\documentclass[manuscript]{acmart}
\usepackage{amsmath,amsfonts}
\usepackage{algorithmic}
\usepackage{graphicx}
\usepackage{textcomp}
\usepackage{xcolor}

\usepackage{multirow}
\usepackage{pbox}
\usepackage{xspace}
\usepackage{makecell}
\usepackage[colorinlistoftodos]{todonotes}

\newcommand{\eg}{\textit{e.g.}\@\xspace}
\newcommand{\ie}{\textit{i.e.}\@\xspace}
\newcommand{\etal}{\textit{et al.}}
\newcommand\invertedComma[1]{``#1''}
\usepackage{lineno,hyperref}
\modulolinenumbers[5]
\usepackage{booktabs}
\usepackage{array}
\usepackage{color, colortbl}
\definecolor{Gray}{gray}{0.9}
\usepackage{caption}

\AtBeginDocument{
  \providecommand\BibTeX{{
    \normalfont B\kern-0.5em{\scshape i\kern-0.25em b}\kern-0.8em\TeX}}}   

\AtBeginDocument{
  \providecommand\BibTeX{{
    Bib\TeX}}}

\setcopyright{acmcopyright}
\copyrightyear{2023}
\acmYear{2023}
\acmDOI{XXXXXXX.XXXXXXX}

\acmConference[CHI '23 CCAI Workshop]{ACM CHI Workshop on
Child-Centred AI Design: Definition, Operation, and Considerations}{April 23--28,
  2023}{Hamburg, Germany}

\acmPrice{15.00}
\acmISBN{978-1-4503-XXXX-X/18/06}

\begin{document}

\title{Examining the Values Reflected by Children during AI Problem Formulation}

\author{Utkarsh Dwivedi, Salma Elsayed-Ali, Elizabeth Bonsignore \&  Hernisa Kacorri}
\email{(udwivedi,sea,ebongin,hernisa)@umd.edu}
\affiliation{
  \institution{\; University of Maryland, College Park}
  \city{College Park}
  \state{Maryland}
  \country{USA}
}

\renewcommand{\shortauthors}{Dwivedi et al.}

\begin{abstract}
Understanding how children design and what they value in AI interfaces that allow them to explicitly train their models such as teachable machines, could help increase such activities' impact and guide the design of future technologies.  In a co-design session using a modified storyboard, a team of 5 children (aged 7-13 years) and adult co-designers, engaged in AI problem formulation activities where they imagine their own teachable machines. Our findings, leveraging an established psychological value framework (the Rokeach Value Survey), illuminate how children conceptualize and embed their values in AI systems that they themselves devise to support their everyday activities. Specifically, we find that children's proposed ideas require advanced system intelligence, \eg emotion detection and understanding the social relationships of a user. The underlying models could be trained under multiple modalities and any errors would be fixed by adding more data or by anticipating negative examples. Children's ideas showed they cared about family and expected machines to understand their social context before making decisions.
\end{abstract}


\maketitle

\section{Introduction}
\begin{figure*}
\centering

\includegraphics[width=.7\textwidth]{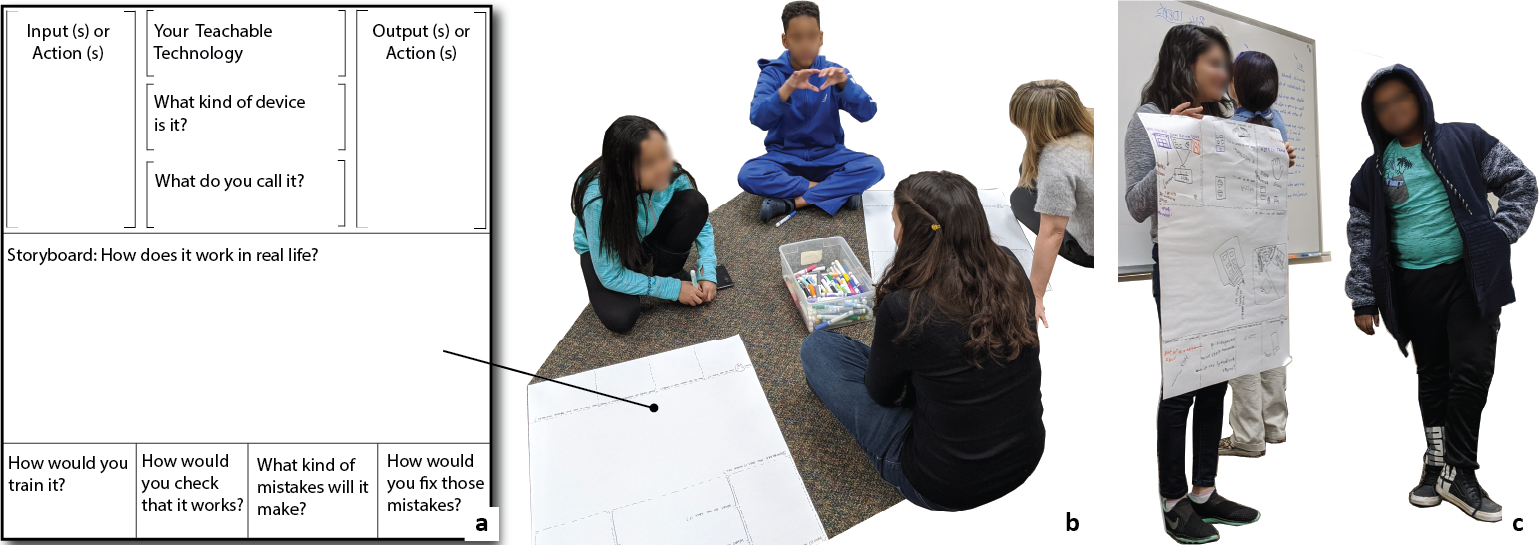}

\captionof{figure}{Our co-design study comprises: a) Modified \textit{Big Paper} with a structured storyboard that b) child-adult pairs use to frame and discuss their ideas and then c) present their teachable machines while a researcher summarizes their input.}
 \label{fig:teaser}
  \end{figure*}

Gaining insight into children's values is a longstanding challenge in Child-Computer Interaction (CCI)~\cite{elsayed-ali2020designing,spiel2018micro-ethics,van_mechelen2014applying,skovbjerg2016being,druin2002role,iversen2012values-led,iversen2012working}. Aligned with these efforts, in this work we involve children as active participants in designing future technology to reveal how they conceptualize and embed their values in a variety of AI system that they envision will support their everyday lives. The technology children are called to design is a teachable machine\footnote{A term first coined in 1978~\cite{andreae1978teachable} and more recently used by Google~\cite{google2017teachable}.},
where ``\textit{the user is a willing participant in the adaptation process and actively provides feedback to the machine to guide its learning}''~\cite{patel1998teachable}. Teachable machines are playing an essential role in designing activities that can spark curiosity and promote children’s understanding of how AI works, its limitations, how bias creeps in, and its potential implications for our society~\cite{dwivedi2021introducing, vartiainen2021machine, hitron2019can}. To elicit children's interests and values we don't ask them to tinker with the input or output of existing teachable machines \eg, recognize objects~\cite{vartiainen2020learning, dwivedi2021introducing} or respond to gestures~\cite{hitron2019can}. Instead, we invite them to imagine their own. Specifically, we invite a university-based intergenerational design team of 5 children (aged 8-13 years) and 5 adults, who have previously engaged in typical black-box tinkering teachable machines activities (\eg, similar to Dwivedi~\etal~\cite{dwivedi2021introducing} and Vartiainen~\etal~\cite{vartiainen2021machine}). 
As shown in Figure~\ref{fig:teaser}, we employ a structured ``Big Paper'' technique~\cite{fails2012methods}. 
Structured storyboards allow us to explore how children define the context (\ie everyday use cases) as well as input and outputs for their personal AI-infused technologies. More so, they explicitly elicit children's notions of potential system failures their machine might encounter, along with their proposed approach for error recovery.

We used the Rokeach Value Survey~\cite{rokeach1973nature}, an established psychological value framework, as an analytical framework to examine the values reflected in children's designs while triangulating data from 5 \invertedComma{Big Paper} storyboards, 2 hours of session recordings, and 2 sets of field notes.
We contribute evidence that participatory machine learning approaches that engage children in AI problem formulation can offer designers and researchers a means for re-centering children in the design of AI-infused technologies and for gaining insights into children's values in AI.
Specifically, our exploratory study: (i) provides empirical results on how children (8-13 years old) formulate the everyday issues their AI systems will address; and
(ii) provides an approach for inferring human values in participatory machine learning projects by using Rokeach's value survey~\cite{rokeach1973nature} as an analytical lens for interpreting the personal values in children's designs.

\section{Related Work}
Here, we summarize two strands of related work that supported our efforts to co-design the everyday AI systems that children envisioned.

\subsection{Problem Formulation via Participatory Machine Learning}
There is a recent push for participatory techniques in machine learning (see Sloane \etal~\cite{sloane2020participation} for a detailed review). 
The term `participatory machine learning' is often used when participatory techniques are included in the problem formulation phase (\eg,~\cite{fernanda2021participatory, martin2020participatory}). 
Martin \etal~\cite{martin2020participatory} state that problem formulation, the focus of our co-design work, is a vital step in any machine learning application. 
They define it as\textit{\invertedComma{a step that involves determining the strategic goals driving the interventions and translating those strategic goals into tractable machine learning problems.}} 
Practical guidelines have emerged to help developers formulate their problems in machine learning~\cite{google2022formulate, amazon2022formulating}. 
However, these guidelines typically do not mention or recommend stakeholder involvement.
Most prior work refers to adults as stakeholders; very few studies have engaged children in problem formulation via participatory machine learning (\eg ~\cite{woodward2018using, vartiainen2021machine, druga2022family}).

\subsection{Children's Values in Design}
Technology is inherently value-laden~\cite{eriksson2022teaching, verbeek2011moralizing, iversen2012values-led, van_mechelen2014applying} 
and computing research on values has largely focused on values of ethical import~\cite{muller1997toward,friedman1996value-sensitive} such as privacy and security. 
Few of these studies focus on  \textit{personal values} and they largely emphasize the values of adults and researchers rather than children~\cite{skovbjerg2016being,yarosh2011examining,kawas2020another,van_mechelen2014applying}. 
Skovbjerg \etal~\cite{skovbjerg2016being} recommend that Child-Computer Interaction (CCI) researchers acknowledge and discuss the embedded values in systems from the outset. In case of AI, important initial design implications can emerge in the problem formulation phase, raising profoundly different ethical concerns, such as possible threats to fairness and civil rights ~\cite{passi2019problem}. 
Thus, it is critical to examine and explicitly foreground children’s personal values in the design of AI-infused technology, and technology more broadly~\cite{elsayed-ali2020designing,druin2002role,yarosh2011examining,skovbjerg2016being,spiel2018micro-ethics}. 
Social psychologist Milton Rokeach defines values as enduring beliefs and personal standards that guide and determine actions, attitudes, ideologies, judgments, justifications, and presentations of self~\cite{rokeach1973nature}. 
In his established value survey (RVS), he identified 36 values: 18 instrumental reflecting preferred behaviors and 18 terminal reflecting preferred end-states of existence~\cite{rokeach1973nature,iversen2012values-led,voida2005conveying,elsayed-ali2020designing}. 
We use these as an analytical tool to examine the perceived values reflected in children's designs. 

\section{Insights}
We developed an understanding of the instrumental and terminal values in Rokeach's framework~\cite{rokeach1973nature} from past work by He \etal~\cite{he2010one} who used the RVS to classify persuasive energy feedback technologies. Similarly, we sought to employ the RVS as an instrument for analyzing children's values. Using Rokeach's 36 values and their definitions, two coders found patterns in the data from 5 \invertedComma{Big Paper} storyboards, 2 hours of session recordings, and 2 sets of field notes. They individually coded the data, discussed classifications, resolved disagreements, and reached consensus. 

\subsection{Instrumental Values in Children's Designs}
Rokeach describes instrumental values as \textit{preferred modes of behavior}. For example, the value of ``capability'' entails competency and efficiency. Of the 18 instrumental values, we observed 5 in children's designs of their everyday AI systems.
All of the children's designs appealed to the following instrumental values: ``capability,'' ``logic,'' ``helpfulness,'' and ``responsibility.''
Children described how their teachable machines would assist them in accomplishing a specific task, often explicitly using the word ``help'' to describe their machines. For example, Penny (a pseudonym) stated, \textit{``My machine is a wearable sensor, like a bracelet, and it’s a bracelet for cheerleaders to \textbf{help} them stay safe.''} Not only would Penny's safety bracelet ensure the welfare of the wearer, but also anyone within a close distance to the wearer. She continued, \textit{``it tells people to get off the mat because it measures the people who are on the mat first and it alerts them to get off the mat to back up, and it measures if someone is still on or not.''}

\subsection{Terminal Values in Children's Designs}
Rokeach describes terminal values as \textit{preferred end-states}.
For example, the value of ``family security'' would entail taking care of loved ones.
Of the 18 values, we observed 4 that were present in children's designs.
We classify children's designs as appealing to the following terminal values: ``family security,'' ``a comfortable life,'' ``inner harmony,'' and ``an exciting life.'' For example, Alan's teachable machine reflected Rokeach's value ``inner harmony,'' which means to have freedom from inner conflict. Alan designed an emotion detection machine in order to identify and alleviate times when someone may be experiencing negative emotions like sadness or anger, as well as detect and reward times when someone is happy or excited.

\section{Conclusion}
Our exploratory co-design session offers insights into the everyday issues children imagine that AI systems can support (i.e., children's AI problem formulation). Moreover, we 
provide an approach for inferring human values in participatory machine learning projects by using Rokeach’s value survey [22] as an analytical lens for interpreting the personal values in children’s designs. We hope to engage other workshop participants in future collaborations that may benefit from these insights, or provide constructive critiques of their potential utility in child-centered AI system design.
\bibliographystyle{ACM-Reference-Format}
\bibliography{sample-base}
\end{document}